\documentclass[aps,twocolumn,prd,showpacs,nofootinbib]{revtex4}
\usepackage{amsmath}
\usepackage{graphicx}
\usepackage{dcolumn}
\usepackage{bm}
\usepackage{amssymb}
\usepackage{latexsym}

\begin{document}

\title[Casimir Effect in Compact Spaces]
{Casimir Effect for Gauge Fields in Spaces with Negative
Constant Curvature}

\author{A. A. Bytsenko}
\address{Departamento de F\'{\i}sica, Universidade Estadual de Londrina,
Caixa Postal 6001, Londrina-Paran\'a, Brazil \\
{\em E-mail address:}
{\rm abyts@uel.br}}

\author{M. E. X. Guimar\~aes}
\address{Departamento de
Matem\'atica, Universidade de Bras\'{\i}lia, Bras\'{\i}lia, DF, Brazil \\
{\em E-mail address:} {\rm marg@unb.br}}

\author{V. S. Mendes}
\address{Departamento de F\'{\i}sica, Universidade Estadual de Londrina,
Caixa Postal 6001, Londrina-Paran\'a, Brazil \\
{\em E-mail address:}
{\rm vsmendes@yahoo.com.br}}

\date{September, 2004}

\thanks{}

\bigskip

\begin{abstract}

We consider gauge theories based on abelian $p-$forms on real
compact hyperbolic spaces. Using the zeta-function regularization
method and the trace tensor kernel formula, we determine
explicitly an expression for the vacuum energy (Casimir energy)
corresponding to skew-symmetric tensor fields. It is shown that
the topological component of the Casimir energy for co-exact forms
on even-dimensional spaces, associated with the trivial character,
is always negative. We infer on the possible cosmological
consequences of this result.
\end{abstract}
\pacs{04.70.Dy, 11.25.Mj}

\maketitle

\section{Introduction}

The topological Casimir effect for scalar (or spinor)
fields on spaces of form $\Gamma\backslash {\widetilde X}$, where
$\Gamma$ is a discrete group acting on manifold ${\widetilde X}$,
has become a very exciting and important issue in areas of
quantum field theory and cosmology [1-9].
Initial evaluation of the Casimir energy has been given for
${\widetilde X} = {\mathbb R}^N$, ${\mathbb S}^N$. In [7-15], the
calculation involves the case in which ${\widetilde X}$
is a Lobachevsky real hyperbolic space.

Maximally symmetric spaces, such as the hyperbolic spaces, play
very important in supergravity \cite{sugra}, in superstring theory
\cite{string} and definitely plays a crucial role in cosmology
\cite{frw, frank, kaloper}. Besides, hyperbolic space forms are
examples of a general noncompact irreducible rank one symmetric
space the proper and outlook mathematical machinery could be
available. In this paper we present a decomposition of the Hodge
Laplacian and the tensor kernel trace formula for free generalized
gauge fields ($p-$forms) on real hyperbolic space forms. The main
ingredient required is a type of differential form structure on
the physical, auxiliary, or ghost variables. We evaluate spectral
functions and the Casimir effect associated with physical degrees
of freedom of the Hodge--de Rham operators on $p-$forms. Let
$\omega_p,\, \varphi_p$ be exterior differential $p-$forms; then,
the invariant inner product is defined by $ (\omega_p,
\varphi_p)\stackrel{def}{=}\int_{\tilde X}
\omega_p\wedge*\varphi_p. $ Under the action of the Hodge $*$
operator the following properties for forms hold: $ \omega_p  =
(-1)^{p(n-p)}\omega_p, $ and $ dd = \delta\delta=0,\, \delta =
(-1)^{np+n+1}*d*. $ The operators $d$ and $\delta$ are adjoint to
each other with respect to this inner product for $p-$forms:
$(\delta\omega_p, \varphi_p) = (\omega_p, d\varphi_p)$. In quantum
field theory the Lagrangian associated with $\omega_p$ takes the
form: $ d\omega_p\wedge *d\omega_p\,\,\,\,\, ({\rm gauge\,\,\,\,\,
field})\,,\, $ $ \delta\omega_p\wedge*\delta\omega_p\,\,\,\,\,
({\rm co-gauge\,\,\,\,\, field}). $ The Euler-Lagrange equations,
supplied with the gauge, give: $ {\frak L}_p\omega_p
=0\,,\,\,\delta\omega_p =0 $ (Lorentz gauge); $ {\frak
L}_p\omega_p =0\,,\,\, d\omega_p =0 $ (co-Lorentz gauge). These
Lagrangians give possible representation of tensor fields or
generalized abelian gauge fields.

As an application, we evaluate the Casimir effect associated with
topologically inequivalent configurations of Abelian co-exact
forms on real compact hyperbolic manifolds.

\section{The trace formula applied to the tensor kernel}

Let us consider an $N-$dimensional compact real hyperbolic
space $X_{\Gamma}$
with universal covering $\widetilde{X}$ and fundamental group
$\Gamma$. We can
represent $\widetilde{X}$ as the symmetric space $G/K$,
where $G=SO_1(N,1)$ and
$K=SO(N)$ is a maximal compact subgroup of $G$. Then we regard
$\Gamma$
as a discrete subgroup of $G$ acting isometrically on
$\widetilde{X}$, and we take
$X_{\Gamma}$ to be the quotient space by that action:
$X_{\Gamma}=\Gamma\backslash \widetilde{X} = \Gamma\backslash G/K$.
Let ${\rm Vol}(\Gamma\backslash G)$ denote the integral of the
constant function ${\mathbb I}$ on $\Gamma\backslash G$ with
respect to the $G-$invariant measure on $\Gamma\backslash G$
induced by $dx$,
$
{\rm Vol}\left( \Gamma \backslash G\right) =
\int_{\Gamma \backslash G}{\mathbb I}dx.
$
For $0\leq p\leq N-1$ the Fried trace formula
applied to the heat kernel ${\mathcal K}_t= e^{-t{\frak L}_p}$
holds \cite{Fried}:
\begin{eqnarray}
{\rm Tr}\left(e^{-t{\frak L}_{p}}\right) & = &
I_{\Gamma}^{(p)}({\mathcal K}_t)
+I_{\Gamma}^{(p-1)}({\mathcal K}_t)
\nonumber \\
& + &
H_{\Gamma}^{(p)}({\mathcal K}_t)+
H_{\Gamma}^{(p-1)}({\mathcal K}_t)
\mbox{,}
\label{Fried}
\end{eqnarray}
where $I_{\Gamma}^{(p)}({\mathcal K}_t)\,,
H_{\Gamma}^{(p)}({\mathcal K}_t)$ are the identity and hyperbolic
orbital integrals respectively:
\begin{eqnarray}
I_{\Gamma}^{(p)}({\mathcal K}_t) & \stackrel{def}{=} &
\frac{\chi(1){\rm Vol} (\Gamma\backslash G)}{4\pi}
\nonumber \\
& \times &
\int_{\Bbb R}\mu_{\sigma_p}(r)e^{-t(r^2+p+\rho_0^2)}dr
\mbox{,}
\label{Id}
\end{eqnarray}
\begin{eqnarray}
H_{\Gamma}^{(p)}({\mathcal K}_t) & \stackrel{def}{=} &
\frac{1}{\sqrt{4\pi t}}  \sum_{\gamma\in C_
\Gamma-\{1\}}\frac{\chi(\gamma)}{j(\gamma)}t_\gamma C(\gamma)
\chi_{\sigma_p} (m_\gamma)
\nonumber \\
& \times &
e^{-t(\rho_0^2+p)-
\frac{t_\gamma^2}{4t}}
\mbox{,}
\end{eqnarray}
$C_{\Gamma}\subset\Gamma$ is a complete set of representations in
$\Gamma$ of its conjugacy classes, and $C(\gamma)$ is a well
defined function on $\Gamma - \{1\}$,
$\rho_0=(N-1)/2$, and $\chi_\sigma(m)={\rm trace} (\sigma(m))$
is the character $\sigma$ for $m\in SO(N)$.
The trace formula involves the Harish-Chandra-Plancherel measure
$\mu_{\sigma_p}(r)$ which is given by
\begin{equation}
\mu_{\sigma_p}(r)=
\left(
\begin{array}{c}
N-1 \\
p
\end{array}
\right)
\times
\left\{
\begin{array}{l}
C_{G}\pi P\left( r\right) r\tanh \left( \pi r\right) , \,
\\
\,\,\,\,\,\,\,\,\,\,\,\,\,\,\,\,\,\,\,\,\,\,\,\,\,
\,\,\,\,\,\,
\mbox{for } N=2n
\\
C_{G}\pi P\left( r\right) ,\,\, \mbox{for }N=2n+1
\end{array}
\right.
\label{10}
\end{equation}
where
$
C_{G}=\left(2^{2N-4}\Gamma (N/2)^2\right),
$
and $P\left( r\right) $ is a polynomial, which presents
the following form
\begin{equation}
P\left( r\right) =\left\{
\begin{array}{l}
\prod_{j=0}^{n-2}\left[r^{2}+((2j+1)/2)^{2}\right] =
\sum_{\ell=0}^{n-1}a_{2\ell}r^{2\ell} ,
\\
\,\,\,\,\,\,\,\,\,\,\,\,\,\,\,\,\,\,\,\,\,\,\,\,\,\,\,
\,\,\,\,\,\,\,\,\,\,\,\,\,\,\,\,\,\,\,\,\,\,\,\,\,\,\,
\,\,\,\,\,\,\,\,\,\,\,\,\,\,\,\,\,\,\,\,\,\,
{\rm for}\, N=2n
\\
\
\!\!\!
\prod_{j=0}^{n-1}\left[ r^{2}+j^{2}\right] =
\sum_{\ell=0}^{n}a_{2\ell}r^{2\ell}\!,\, {\rm for}\,N=2n+1
\end{array}
\right.
\label{13}
\end{equation}
$a_{2\ell}$ are the Miatello coefficients
\cite{Miatello,BytsenkoM}.
For $p\geq 1$ there is a measure $\mu_{\sigma}(r)$ corresponding
to a general irreducible representation $\sigma$.
Let $\sigma_p$ be the standard representation of $SO(N-1)$ on
$\Lambda^p{\Bbb C}^{(N-1)}$. If $N=2n$ is even then
$\sigma_p\,\,(0\leq p\leq n-1)$ is always irreducible; if $N=2n+1$
then every $\sigma_p$ is irreducible except for $p=(N-1)/2=n$, in
which case $\sigma_n$ is the direct sum of two spin--$(1/2)$
representations
$\sigma^{\pm}:\,\,\sigma_n=\sigma^{+}\oplus\sigma^{-}$. For $p=n$ the
representation $\tau_n$ of $K=SO(2n)$ on $\Lambda^n {\Bbb C}^{2n}$
is not irreducible: $\tau_n=\tau_n^{+}\oplus\tau_n^{-}$ is the
direct sum of two spin--$(1/2)$ representations.

In the case of the trivial representation ($p=0$, i.e.
for smooth functions or
smooth vector bundle sections) the measure $\mu (r)\equiv
\mu_{0}(r)$ corresponds to the trivial representation. Therefore,
we take $I_{\Gamma}^{(-1)}({\mathcal K}_t)
=H_{\Gamma}^{(-1)}({\mathcal K}_t)=0$. Since
$\sigma_0$ is the trivial representation, one has
$\chi_{\sigma_0}(m_{\gamma})=1$. In this case, formula (\ref{Fried})
reduces exactly to the trace formula for $p=0$
\cite{Bytsenko1,Bytsenko2,Wallach,BytsenkoW}.

\section{The spectral functions of exterior forms and the vacuum energy}

If ${\mathfrak L}_p$ is a self-adjoint Laplacian on $p-$forms then
the following results hold.
There exists $\varepsilon,\delta >0$
such that for $0<t<\delta$ the heat kernel expansion for
Laplace operators on a compact manifold $X_{\Gamma}$ is given by
$
{\rm Tr}\left(e^{-t{\mathfrak L}_p}\right)=
\sum_{0\leq \ell\leq \ell_0} a_\ell
({\mathfrak L}_p)t^{-\ell}+ {\mathcal O}(t^\varepsilon).
$
The zeta function of ${\mathfrak L}_p$ is the Mellin transform
$
\zeta(s|{\mathfrak L}_p)=
(\Gamma(s))^{-1}\int_{{\Bbb R}_{+}}
{\rm Tr}e^{-t{\mathfrak L}_p}t^{s-1}dt
$.
This function equals ${\rm Tr}\left({\mathfrak L}_p^{-s}\right)$ for
$s> (1/2){\rm dim}\,(\Gamma\backslash G)$.

The transverse part of the skew-symmetric tensor is represented by
the co-exact $p-$form $\omega_p^{(CE)}=\delta\omega_{p+1}$, which
trivially satisfies $\delta\omega_p^{(CE)}=0$, and we denote by
${\frak L}_p^{(CE)}$ the restriction of the Laplacian on
the co--exact $p-$form. The goal now is to extract the co--exact
$p-$form on the manifold which describes the physical degrees of
freedom of the system, and presents by alternating sum of forms
\cite{BytsenkoX1,BytsenkoX2,BytsenkoX3}.

\subsection{The identity component of the isometry group}

The zeta function related to the identity integral
$I_{\Gamma}^{(p)}({\mathcal K}_t)$ in (\ref{Id}) has the form
\begin{eqnarray}
\zeta _{I}( s|{\frak L}_p) & = & \frac{\chi \left( 1\right) {\rm
Vol}\left( \Gamma \backslash G\right) } {4\pi \Gamma(s)}
\int_{0}^{\infty }t^{s-1}dt
\nonumber \\
& \times & \int_{\mathbb{R}} \mu_{\sigma_p}(r)e^{-ty(r^2; m^2_p)}dr
\nonumber \\
& = & \frac{1}{4}\chi \left( 1\right) {\rm Vol}\left( \Gamma
\backslash G\right) \nonumber
\\
& \times & \int_{\mathbb{R}}\mu_{\sigma_p}(r)[y(r^2; m^2_p)]^{-s}dr
\mbox{,}
\label{21}
\end{eqnarray}
where $y(r^2; m^2_p)\equiv r^2+m^2_p$,\,
$m^2_p\equiv b^{(p)}+(\rho_0-p)^2$.
Because of Eqs. (\ref{10}) and (\ref{13}) it is convenient to
consider even- and odd-dimensional cases separately.

\subsubsection{Even-dimensional manifold}

Using Eqs. (\ref{10}) and (\ref{13}) in (\ref{21}) for $N=2n$
we get
\begin{eqnarray}
\zeta _{I}^{( 2n)}( s|{\frak L}_p) & = & \frac{1}{4}
\chi(1){\rm Vol}\left( \Gamma \backslash G\right)
C_{G} \int_{\mathbb{R}}\frac{P(r)r\tanh (\pi r)dr}{[y(r^2; m^2_p)]^{s}}
\nonumber \\
& = & \frac{1}{4}\chi(1) {\rm Vol}\left( \Gamma \backslash
G\right) C_{G} \nonumber
\\
& \times & \sum_{j=0}^{n-1}a_{2j}\int_{\mathbb{R}} \frac{r^{2j+1}\tanh
\left( \pi r\right)dr} {[y(r^2; m^2_p)]^{s}} \mbox{.} \label{26}
\end{eqnarray}
For $a, \delta>0$, $z\in {\mathbb C}$, define the entire functions
$
{\frak K}_{m} (s;\delta ,a) \stackrel{def}{=}
\int_{\mathbb{R}} r^{2m}(\delta + r^2)^{-s}
{\rm sech}^{2} ( ar)dr.
$
Then for ${\rm Re}\,s>j+1$, $j\geq 0$, one gets
\cite{BytsenkoW}
\begin{eqnarray}
& & \int_{\mathbb{R}}\frac{r^{2j+1}\tanh \left( ar\right)dr
}{\left( \delta +r^{2}\right) ^{s}}= \nonumber
\\
& & = \frac{aj!}{2}\sum_{\ell=0}^{j} \frac{{\frak K}_{j-1}\left(
s-\ell-1;\delta ,a\right) }{\left(j-\ell \right)! \left(
s-1\right) \left( s-2\right) ...\left( s-\left( \ell+1\right)
\right)},
\label{27}
\end{eqnarray}
The following result follows:
\begin{eqnarray}
& & \zeta _{I}^{(2n)}( s|{\frak L}_p)  = \frac{\pi}{8} \chi(1)
{\rm Vol}(\Gamma \backslash G) C_{\Gamma} \left(
\begin{array}{c}
2n-1 \\
p
\end{array}
\right)\sum_{j=0}^{n-1}a_{2j}j!  \notag \\
& \times & \sum_{\ell=0}^{j}\frac{{\frak K}_{j-1} \left(
s-\ell-1;b^{\left( p\right) }+\left( \rho _{0}-p\right) ^{2},\pi
\right) } {\left( j-\ell\right)! \left( s-1\right) \left(
s-2\right) ... \left( s-\left( \ell+1\right) \right) }.
\label{zeta-i even res}
\end{eqnarray}

\subsubsection{Odd-dimensional manifold}

In odd-dimensional case, $N=2n+1$, we get
\begin{eqnarray}
\zeta _{I}^{\left( 2n+1\right) }(s|{\frak L}_p) & = &
\frac{1}{4}\chi(1){\rm Vol} \left( \Gamma \backslash G\right)
C_{\Gamma} \left(
\begin{array}{c}
2n \\
p
\end{array}
\right) \nonumber \\
& \times & \sum_{j=0}^{n}a_{2j}\int_{\mathbb{R}} r^{2j}[y(r^2;
m^2_p)]^{-s}dr
\mbox{.}
\label{30}
\end{eqnarray}
Using the formula
$
\int_{\mathbb{R}} r^{2j+1}(\delta^2 +r^{2})^{-s}dr=
\delta^{2j-2s}\Gamma(j)\Gamma( s-j)/(\Gamma(s)),
$
we find:
\begin{eqnarray}
& & \zeta _{I}^{( 2n+1)}( s|{\frak L}_p) = \frac{\chi \left(
1\right) {\rm Vol}\left( \Gamma \backslash G\right)
}{4\Gamma(s)}\left(
\begin{array}{c}
2n \\
p
\end{array}
\right)
\nonumber \\
& \times & \sum_{j=0}^{n}\frac{a_{2j}\Gamma \left( j+\frac{1}{2} \right)
\Gamma \left( s-j-\frac{1}{2}\right)} {m_p^{2s-2j-1}} \mbox{.}
\label{odd}
\end{eqnarray}

\subsection{The hyperbolic component of the isometry group}

The zeta function associated with hyperbolic orbital integral
$H_{\Gamma}^{(p)}({\mathcal K}_t)$ takes the form
\begin{eqnarray}
\zeta _{H}(s|{\frak L}_p)  &= &  \frac{1}{\sqrt{4\pi}\Gamma(s)}
\sum_{\gamma \in C_{\Gamma }-\left\{ 1\right\} }\chi \left(
\gamma \right) j^{-1}\left( \gamma \right) t_{\gamma }C \left(
\gamma \right)
\nonumber \\
& \times &
\int_{0}^{\infty }t^{s-\frac{3}{2}}
e^{-tm^2_p+\frac{t_{\gamma}^2}{4t}}dt
\mbox{.}
\label{zeta-h def}
\end{eqnarray}
Using the $K-$Bessel function $K_{\nu}(s),\, s\in {\mathbb C}$,
defined by $ K_{\nu}(2s) \stackrel{def}{=} (1/2)s^{\nu}
\times\int_0^{\infty} t^{-\nu-1}\exp\{-t-s^2/t\}dt, $ we have
\begin{eqnarray}
\zeta _{H}(s|{\frak L}_p) & = &
\frac{1}{\sqrt{\pi }\Gamma \left(s\right)}
\sum_{\gamma \in C_{\Gamma }-\left\{ 1\right\} } \chi \left(
\gamma \right) j^{-1}\left( \gamma \right) C\left(\gamma\right)
\nonumber \\
& \times &
t_{\gamma }^{s+\frac{1}{2}}
\frac{K_{-s+\frac{1}{2}}(t_{\gamma}m_p)} {[ 2m_p]^{s-\frac{1}{2}}}
\nonumber \\
& = & \int_{0}^{\infty }\sum_{\gamma \in C_{\Gamma }-\left\{
1\right\}}\frac{\chi \left( \gamma \right)t_{\gamma}j^{-1} \left(
\gamma \right) C\left( \gamma \right)}{\Gamma(s)\Gamma(1-s)}
\nonumber \\
& \times &
\frac{e^{-(t+m_p)t_{\gamma }}dt} {( 2tm_p+t^2)^{s}}
\mbox{.}
\end{eqnarray}

\subsubsection{Logarithmic derivative of the Selberg zeta function}

The function $\psi_{\Gamma}(s; \chi)$ defined in \cite{Gangolli}
\begin{equation}
\psi_{\Gamma }\left( z;\chi \right) \stackrel{def}{=}
\sum_{\gamma \in C_{\Gamma }-\left\{
1\right\} }\chi \left( \gamma \right) t_{\gamma }j^{-1}
\left( \gamma \right)
C\left( \gamma \right) e^{-\left( z-\rho _{0}\right) t_{\gamma }}
\mbox{,}
\end{equation}
for ${\rm Re}\,s>2\rho_0$, is a holomorphic
function in the half-plane ${\rm Re}\,s>2\rho_0$ and admits
a meromorphic continuation to the full complex plane. It has been
shown that there is a meromorphic function $Z_{\Gamma}(s; \chi)$
on ${\mathbb C}$ such that $(d/dz){\rm log}Z_{\Gamma}(z; \chi) =
\psi_{\Gamma}(z;\chi)$. $Z_{\Gamma}(z; \chi)$ suitable normalized
is the Selberg zeta function attached to $(G,K,\Gamma,\chi)$.
Therefore,
\begin{equation}
\zeta_{H}(s|{\frak L}_p) = \frac{1}{\Gamma \left( s\right)
\Gamma \left( 1-s\right)}\int_{0}^{\infty }\frac{\psi_{\Gamma}
\left( \rho _{0}+t+m_p;\chi \right)}
{(2tm_p+t^{2})^{s}}dt
\mbox{.}
\label{35}
\end{equation}
Canonical quantization of Abelian $p-$forms yields a formal
expression
$(1/2)\zeta(s=-1/2|{\frak L}_p)=(1/2)\sum_{j}\lambda^{1/2}$
for the Casimir energy, where $\{\lambda\}_{j\geq 0}^{\infty}$
is the set of eigenvalues (with multiplicity) of the Laplacian
${\frak L}_p$ on smooth sections of a vector bundle over
$X_{\Gamma} = \Gamma\backslash {\mathbb H}^N$ induced by a
finite-dimensional unitary representation $\chi$ of $\Gamma$.
The regularized Casimir energy related to
co-exact forms (the alternating sum of zero- and $p-$form
components) on real compact even-dimensional hyperbolic
manifolds is given as follows:
\begin{eqnarray}
E(m_p)_{N=2n} & = & \frac{1}{2}\zeta(-1/2|{\frak L}_p^{(CE)})
= \frac{\pi}{16}\chi(1) {\rm Vol}(\Gamma \backslash G)C_{G}
\nonumber \\
& \times &
\sum_{j=0}^{n-1} \sum_{\ell=0}^{j}
\frac{a_{2j}j!}{( j-\ell)!\prod_{q=0}^{\ell}(-\frac{1}{2}-
\left(q +1\right))}
\nonumber \\
& \times &
\left[
\left(
\begin{array}{c}
2n-1 \\
p
\end{array}
\right)
{\frak K}_{j-1}( -\ell-\frac{3}{2}; m^2_p,\pi)
\right.
\nonumber \\
& + &
\left.
{\frak K}_{j-1}( -\ell-\frac{3}{2}; m^2_0,\pi)\right]
\nonumber \\
& - & \frac{1}{2\pi}\int_0^{\infty}
\left\{\psi_{\Gamma}(\rho_0+t+m_p; \chi)[2tm_p+t^2]^{\frac{1}{2}}
\right.
\nonumber \\
& + &
\left. \psi_{\Gamma}(\rho_0+t+m_0; \chi)[2tm_0+t^2]^{\frac{1}{2}}
\right\}dt
\mbox{.}
\label{vacuum}
\end{eqnarray}
In the case of scalar field ($p=0$ for a trivial representation)
the Casimir energy becomes
\begin{eqnarray}
&& E(m_0)_{N=2n} =
\frac{\pi}{16}\chi(1) {\rm Vol}(\Gamma \backslash G)C_{G}
\nonumber \\
&& \times
\sum_{j=0}^{n-1} \sum_{\ell=0}^{j}
\frac{a_{2j}j!{\frak K}_{j-1}( -\ell-\frac{3}{2}; m^2_0,\pi)}
{( j-\ell)!\prod_{q=0}^{\ell}(-\frac{1}{2}-
\left(q +1\right))}
\nonumber \\
&& - \frac{1}{2\pi}\int_0^{\infty}
\psi_{\Gamma}(\rho_0+t+m_0; \chi)[2tm_0+t^2]^{\frac{1}{2}}dt
\mbox{.}
\label{vacuum1}
\end{eqnarray}

Formula (\ref{vacuum}) with positive parameter $m_p$ gives the
regularized vacuum energy $E(m_p)$ which is finite. From
(\ref{odd}) it follows that in the case of odd $N$ the identity
component of $E(m_p)$ has poles at $s =- 1/2$ and therefore
$E(m_p)$ cannot be obtained by the method available for
even-dimensional manifolds, which agrees with result obtained in
\cite{Bytsenko7}. For the trivial representation $\chi =1$ of
$\Gamma$, the topological component of the Casimir energy $E(m_p)$
(the last term in (\ref{vacuum})) is always negative, in agreement
with results previously obtained in \cite{cog}. In the case of
scalar fields (zero-forms) our result agrees with one founded in
\cite{Bytsenko7}.

\section{Concluding remarks}

Cosmological predictions, such as the microwave background
anisotropies (CMB) and the current acceleration expansion of the
universe \cite{perl}, depend pretty much on the details of the
theoretical model under consideration. In particular, the recent
data obtained by the Wilkinson Microwave Anisotropy Probe (WMAP)
\cite{wmap} satellite confirmed, and set new standards of
accuracy, to the previous COBE's measurement of a low quadrupole
moment in the angular power spectrum of the CMB, which is in
accordance with the assumption that the topology of the universe
might be non-trivial, with particular emphasis on the case of a
compact hyperbolic universe. Combined with this observation, the
WMAP satellite also indicates that $\sim 60 \% $ of the critical
energy density of the universe is contributed by a smoothly
distributed vacuum energy (Casimir energy) or dark energy, whose
net effect is repulsive (leading, thus, to an accelerated
expansion of the universe).

In this paper, we have shown that the topological component of the
Casimir energy for co-exact forms on even-dimensional manifolds,
associated with the trivial character, is always negative. This
result confirms the above mentioned measurements and we can infer
on the cosmological consequences of it. We plan to address this
question in details in a forthcoming paper \cite{mendes2}.

\section*{Acknowledgements}

A. A. Bytsenko and M. E. X. Guimar\~aes would like to thank the
Conselho Nacional de Desenvolvimento Cient\'{\i}fico e
Tecnol\'ogico (CNPq/Brazil) for partial support. V. S. Mendes
would like to thank CAPES for a PhD grant.


\vspace{0.1cm}






\begin{thebibliography}{99}



\bibitem{DeWitt}
B. De Witt, {\it Phys. Rep.}{\bf 19}, 295 (1975).

\bibitem{Birrell}
N. Birrell and P. W. Davis, {\it Quantum Fields on Curved Soaces}
(Cambridge University Press, Cambridge, 1982).

\bibitem{Ambjorn}
J. Ambjorn and S. Wolfram, {\it Ann. Phys. (N.Y.)}{\bf 147}, 1 (1983).

\bibitem{Greiner}
W. Greiner, B. Muller and G. Plunien, {\it Phys. Rep.} {\bf 134}, 87 (1986).

\bibitem{Jackson}
A. Jackson and L. Vepstas, {\it Phys. Rep.} {\bf 187}, 109 (1990).

\bibitem{Mostepanenko}
V. Mostepanenko and N. Trunov, {\it Casimir Effects and Its
Applications} (Energoatomizdat, Moscow, 1990).

\bibitem{Bytsenko1}
E. Elizalde, S. D. Odintsov, A. Romeo, A. A. Bytsenko and
S. Zerbini, {\it Zeta Regularization Techniques with Applications}
(World Scientific, Singapore, 1994).

\bibitem{Bytsenko2}
A. A. Bytsenko, G. Cognola, L. Vanzo and S. Zerbini,
{\it Phys. Rep.} {\bf 266}, 1 (1996).

\bibitem{Bytsenko3}
A. A. Bytsenko, G. Cognola, E. Elizalde, V. Moretti and
S. Zerbini, {\it Analytic Aspects of Quantum Fields}
(World Scientific, Singapore, 2003).

\bibitem{Bytsenko4}
A. A. Bytsenko and Yu. P. Goncharov, {\it Class. Quantum Grav.}
{\bf 8}, 2269 (1991).

\bibitem{Bytsenko5}
A. A. Bytsenko and Yu. P. Goncharov, {\it Class. Quantum Grav.}
{\bf 8}, L211 (1991).

\bibitem{Bytsenko6}
A. A. Bytsenko and Yu. P. Goncharov, {\it Mod. Phys. Lett. A}
{\bf 6}, 669 (1991).

\bibitem{Bytsenko7}
A. A. Bytsenko and S. Zerbini, {\it Class. Quantum Grav.}
{\bf 9}, 1365 (1992).

\bibitem{cog} G. Cognola, L. Vanzo and S. Zerbini, {\it J. Math.
Phys.} {\bf 32}, 222 (1992).

\bibitem{cog2} G. Cognola, K. Kirsten and S. Zerbini, {\it Phys.
Rev. D} {\bf 48}, 790 (1993).

\bibitem{sugra} M. J. Duff, B. E. Nilsson and C. N. Pope, {\it
Phys. Rep.} {\bf 130}, 1 (1986).

\bibitem{string}
J. Maldacena, {\it Adv. Theor. Math. Phys.} {\bf 2}, 231 (1998).

\bibitem{frw}
E. Elizalde, e-Print arXiv: hep-th/0311195.

\bibitem{frank}
R. Aurich, S. Lustig, F. Steiner and H. Then, e-Print arXiv:
astro-ph/0403597.

\bibitem{kaloper}
N. Kaloper, J. March-Russell, G. D. Starkman and M. Trodden, {\it
Phys. Rev. Lett.} {\bf 85}, 928 (2000).

\bibitem{Fried}
D. Fried, {\it Invent. Math.} {\bf 84}, 523 (1986).

\bibitem{Miatello}
R. Miatello, {\it Trans. Am. Math. Soc.} {\bf 260}, 1 (1980).

\bibitem{BytsenkoM}
A. A. Bytsenko, E. Elizalde and M. E. X. Guimar\~{a}es, {\it Int.
J. Mod. Phys. A} {\bf 18}, 2179 (2003).

\bibitem{Wallach}
N. Wallach, {\it J. Diff. Geom.} {\bf 11}, 91 (1976).

\bibitem{BytsenkoW}
A. A. Bytsenko and F. L. Williams, {\it J. Math. Phys.} {\bf 266}, 1075
(1998).

\bibitem{BytsenkoX1}
A. A. Bytsenko, L. Vanzo and S. Zerbini, {\it Nucl. Phys.}
{\bf B505}, 641 (1997).

\bibitem{BytsenkoX2}
A. A. Bytsenko, {\it Nucl. Phys. (Proc. Suppl.)} {\bf B104}, 127
(2002).

\bibitem{BytsenkoX3}
A. A. Bytsenko, A. E. Gon\c calves and F. L. Williams, {\it Int.
J. Mod. Phys. A} {\bf 18}, 2041 (2003).

\bibitem{Gangolli}
R. Gangolli, {\it Ill. J. Math.} {\bf 21}, 1 (1977).

\bibitem{perl}
S. Perlmutter et al., {\it Ap. J.} {\bf 517}, 565 (1999).

\bibitem{wmap}
H. V. Peiris et al., {\it Ap. JS} {\bf 148}, 213 (2003).


\bibitem{mendes2}
A. A. Bytsenko, M. E. X. Guimar\~aes and V. S. Mendes, {\it in
preparation}.

\end{thebibliography}
\end{document}